# Threshold Voltage-Defined Switches for Programmable Gates

Anirudh Iyengar and Swaroop Ghosh

Computer Science and Engineering, University of South Florida, Tampa, FL-33620

(December 4, 2015)

**Abstract**

Semiconductor supply chain is increasingly getting exposed to variety of security attacks such as Trojan insertion, cloning, counterfeiting, reverse engineering (RE), piracy of Intellectual Property (IP) or Integrated Circuit (IC) and side-channel analysis due to involvement of untrusted parties. In this paper, we propose transistor *threshold voltage-defined switches* to camouflage the logic gate both logically and physically to resist against RE and IP piracy. The proposed gate can function as NAND, AND, NOR, OR, XOR, XNOR, INV and BUF robustly using threshold-defined switches. The camouflaged design operates at nominal voltage and obeys conventional reliability limits. The proposed gate can also be used to personalize the design during manufacturing.

## Introduction

Camouflaging is a technique of hiding the circuit functionality of a few chosen gates to make RE/piracy impossible or extremely hard [1-6]. Camouflaging of gates using dummy contacts [1-2] can realize 3 functions at the cost of *~5X area and power overhead*. Although dummy contacts hide the functionality it requires *process change* (hollow via) and *fails to force exhaustive RE by attacker*. Programmable standard cells using control signals [3] require signal routing for each camouflaged gate. Techniques to deceive the attacker using filler cells [4] and dummy transistors [5] are also proposed. Unlike the proposed threshold voltage ($V_T$) programmable technique *the existing art is either process costly (extra mask costs), leave layout clues (increases design overhead) or offers limited RE resistance*. The proposed camouflaging is complementary to existing obfuscation techniques that hide the functionality of a design by inserting additional components. For sequential circuits, additional logic (black) states are introduced in the finite state machine [7-10], which allow the design to reach a valid state only using the correct key. In combinational logic, XOR/XNOR gates are introduced to conceal the functionality [11-12]. Watermarking and passive metering techniques are also proposed to detect IC piracy [13-14].

The proposed technology addresses the attack model where an adversary can perform invasive RE of a chip to compromise sensitive/classified information, reproduce or sell the pirated copy of the design. The adversary can still create a partial netlist with known gates and RE the missing gate functionalities recursively through carefully selected test patterns. In order to increase the RE difficulty, the camouflaging technique is achieved through $V_T$ modulation (implemented by changing channel doping concentration during manufacturing) of switches which leaves no layout trace. The proposed camouflaged gate can assume 8 functions to obscure the design and force the attacker to launch brute force at least (or nullify the attack in some cases). It must be noted that, camouflaging is associated with area, power and delay overheads. Therefore it is critical to realize the many functionalities with a minimal design overhead.

## Proposed Approach

We propose a novel switch that turns ON/OFF based on threshold voltage ($V_T$) asserted on it. The switch is realized by using conventional NMOS and PMOS transistors with the gate biased at mid-point between nominal N and P threshold voltage i.e., $0.5(V_{TN}+V_{TP})$. Therefore, the switch conducts when low $V_T$ (LVT) is assigned and stops conducting when high $V_T$ (HVT) is assigned during manufacturing (Fig. 1). The proposed switch can be used in conjunction with nominal $V_T$ (NVT) transistors to camouflage the gate. Although the switches are easy to identify in the layout, the $V_T$ of the switch is opaque to the attacker thus, making the configuration secure. The switch configures the functional transistors to serve as NAND, AND, NOR, OR, XOR XNOR, INV or BUF. Thus forcing the adversary to resort to a brute-force attack. It is notable that due to high overheads, the camouflaged gates should be used judiciously. The choice of gates should be guided by metrics that maximize RE effort while lowering overhead and maintaining robustness.

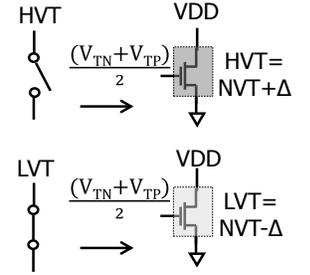

Fig. 1 VT programmable switch. HVT: OFF, LVT: ON. PMOS switch works similarly.

$V_T$ modulation is a well-known technique used extensively in semiconductor industry for trade-off between power, performance and robustness. Therefore, the proposed $V_T$ based camouflaging comes without process cost adder. Since $V_T$ programming can be achieved by channel doping during manufacturing (no layout clues) the RE effort will increase. Due to 8 hidden functions the proposed camouflaging *will require brute force RE at the least*. By camouflaging certain gate sequences (arbitrary gate followed by XOR/XNOR) the design can be *obscured*. Considering 10K design with 50-inputs camouflaging 1% gates will require at least $2^{50}$ RE trials which is $10^5$ years at 1GHz test frequency. Limited usage of camouflaged gates especially in critical paths can keep the timing, area and power impact below 2-3%. Apart from main stream electronics and services, the proposed technology can find applications in military electronics (satellite, radar, guided-missile, unmanned vehicle, rockets etc.), used in mission critical systems and intelligence agencies.

The novelty of this work lies in (i) $V_T$ programmable switch; (ii) camouflaging without any process cost adder; (iii) camouflaged gate that can hide 8 functions; and, (iv) novel attack models such as RE using heating/cooling and side channel analysis.

## Design and Analysis of VT Defined Switch and Camouflaged Gate

There are three aspects of camouflaged gate design: (i) switch; (ii) number of functionalities offered; and, (iii) optimized and robust implementation. The design objective of the switch is to achieve high $I_{ON}/I_{OFF}$ ratio whereas the gate is expected to provide many functionalities at low overhead and high robustness.

### A. Threshold programmable switch design

The switch design is quantified using $I_{ON}/I_{OFF}$ ratio, wherein the gate voltage, HVT, LVT values and transistor sizes are tuned to maximize the $I_{ON}/I_{OFF}$ ratio. For N-switch, higher HVT and lower gate voltage is good for leakage whereas lower LVT and higher gate voltage is good for performance and, vice-versa is true for P-switch (Fig. 3(a)). Initial results using predictive 22nm tech [15] at 1V shows (Fig. 3(b),(c)) that HVT (LVT) target should be ~0.35-0.4V above (below) NVT for best performance. Switch gate voltage can also be tuned to improve speed (Fig. 3(d)).

### B. Camouflaged gate design

Fig. 2(a) shows a conceptual schematic of the camouflaged gate that can assume multiple functionalities. The switches of selected (unselected) function are programmed to LVT (HVT) whereas the function can be implemented using NVT. An example schematic and layout that implements NAND, AND, NOR, OR, XOR and XNOR depending on the $V_T$ of switch 1-10 is shown (Fig. 2(b)-(c)). The ON switches for different functions is also displayed. *The proposed gate ensures brute-force for RE (00/11 is needed to identify XOR/XNOR and 01/10 is needed for*

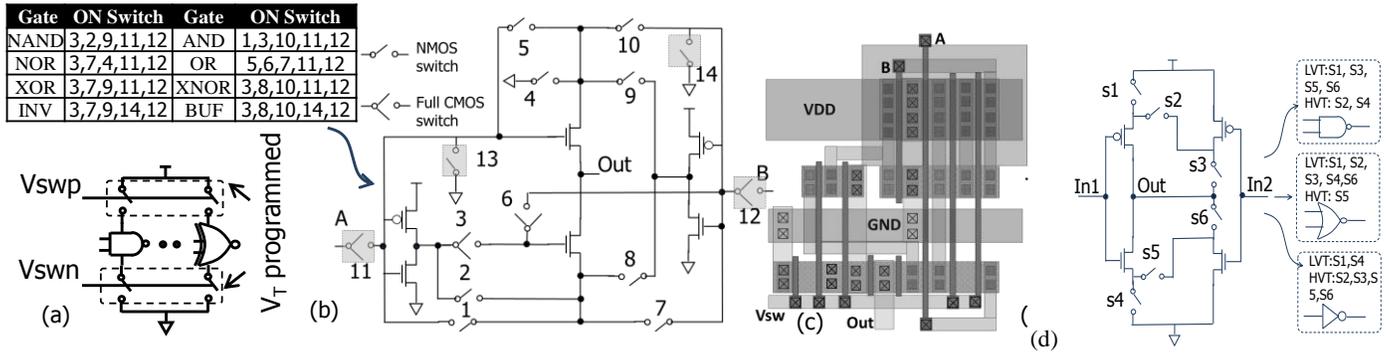

Fig. 2 (a) Conceptual example of camouflaged gate with threshold programmable switch. Only selected gate is connected to power rail; (b) pass transistor-based camouflaged gate to hide 8 functionalities; (c) layout; and, (d) CMOS implementation (supports 3 functions) for risk mitigation.

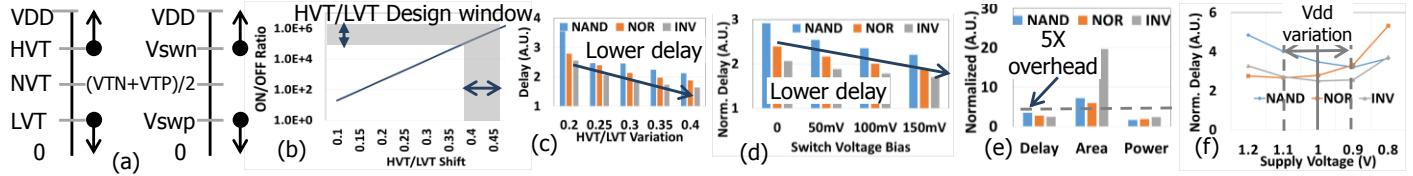

Fig. 3 (a) $V_T$ and switch voltage design window; (b) $V_T$ design window for HVT and LVT of switch for good $I_{ON}/I_{OFF}$ ratio; (c) HVT/LVT window for delay. Mean value of +/-0.25VDD than NVT is used; (d) P and N switch voltage biasing for robustness; (e) design overheads of proposed gate; and (f) impact of supply voltage variation.

NAND/NOR/AND/OR). The gate contains extra switches 11-14 to disconnect the inputs and connect to GND. This feature will fake the inverter (buffer) as a 2-input gate. A redundant signal is connected to 1st input which is floating and the 2nd input is connected to the signal to be inverted (buffered) and, the gate is programmed as XNOR (XOR). Redundant switches can also be added to increase obscurity. The overheads of the proposed gate (Fig. 3(e)) indicate that it should be used judiciously in the design. Simulations on supply voltage variation indicate that both high and low voltage can worsen the delay as it weakens ON switches and strengthens OFF switches to create contention between P and N network. The impact of $V_T$ variation due to temperature or process variation shows similar effect (Fig. 3(f)). The design is optimized to lower delay overhead by (i) tuning the $V_T$ of HVT and LVT transistors; and, (ii) separating the P and N switch gate voltages and biasing them to improve the robustness. Fig. 3(d) shows +/-100mV biasing of switch gate voltage in addition to +/-100mV $V_T$ biasing of HVT and LVT improves the delay by 20%.

The proposed camouflaged gate (Fig. 2(b)) can suffer from area, power and delay overheads. Three options can be pursued for mitigation: (i) simplify the design by eliminating few switches to realize less functions; (ii) considering full CMOS gate structure (Fig. 2(d)); and,(iii) creating two flavors of camouflaged gate to realize {NAND,NOR,XOR} and {AND,OR,XNOR}. To mitigate delay overhead HVT can be used for off-critical path and LVT for critical paths.

## Vulnerabilities, attack models and countermeasures

The leakage and delay of proposed design will change with temperature (due to $V_T$ variation). Adversary can exploit this to perform a side channel analysis to crack the camouflaging. The LVT programmed switches can also be identified using backside probing techniques like LIVA [16].

One model is to heat/cool the chip that will lower/increase the HVT switch $V_T$ and create sneak path from VDD (Fig. 2(d)). The leakage sensitivity of NAND compared to NOR for 2'b11 input will also increase/decrease. Temperature impact on gate delay can be used to obtain different gate delay sensitivities. One possible countermeasure is balancing all flavors of camouflaged gates to contaminate the leakage and delay signature. Another possible countermeasure is to use thermal sensor and dynamically modulate the switch gate voltage to kill the leakage.

## Conclusions

We propose a novel threshold voltage-defined switches to camouflage the logic gate both logically and physically to resist against RE and IP piracy. The proposed transmission gate-based camouflaged gate can assume the role of wide variety of gates including NAND, AND, NOR, OR, XOR, XNOR, INV and BUF robustly using threshold defined switches. The camouflaged design operates at nominal voltage and obeys conventional reliability limits. We also proposed novel attack models such as heating/cooling-assisted RE with side channel analysis.